\RequirePackage{fix-cm}
\documentclass[smallextended]{svjour3}       

\smartqed  
\usepackage{tabu} 
\usepackage{booktabs} 
\usepackage{import}
\usepackage{comment}
\usepackage{graphicx}
\usepackage{paralist}
\usepackage{xspace}
\usepackage{natbib}
\usepackage{caption}
\usepackage{subcaption}
\usepackage{verbatim}
\usepackage[pdftex,colorlinks=true,pdfstartview=FitV,
 	linkcolor=black,citecolor=black,urlcolor=black,bookmarks=false]{hyperref}
\usepackage{mathptmx}      
\usepackage{latexsym}
\newcommand{\eg}{\emph{e.g.},\xspace}
\newcommand{\etal}{\emph{et al.}\xspace}
\usepackage{paralist}
\usepackage{geometry}
\usepackage{amssymb}
\usepackage{ifthen}
\usepackage[normalem]{ulem} 
\usepackage{xcolor}

\newboolean{showedits}
\setboolean{showedits}{true} 
\ifthenelse{\boolean{showedits}}
{
  \newcommand{\del}[1]{\textcolor{red}{\sout{#1}}} 
}{
  \newcommand{\del}[1]{} 
  
}
\newboolean{showcomments}
\setboolean{showcomments}{true}
\newcommand{\id}[1]{$-$Id: scg-llncs.tex 30911 2010-02-05 10:21:47Z oscar $-$}

\ifthenelse{\boolean{showcomments}}
{\newcommand{\nbc}[3]{
 {\colorbox{#3}{\bfseries\sffamily\scriptsize\textcolor{white}{#1}}}
 {\textcolor{#3}{\sf\small$\blacktriangleright$\textit{#2}$\blacktriangleleft$}}}
 }
{\newcommand{\nbc}[3]{}
 \renewcommand{\del}[1]{} 
 }

\usepackage{xspace}
\usepackage{paralist}
\usepackage[inline]{enumitem}

\journalname{Journal of Visualization}
\begin{document}

\title{Bubble Storytelling with Automated Animation:\\{A Brexit Hashtag Activism Case Study} 
}


\author{Noptanit Chotisarn \and
Junhua Lu \and
Libinzi Ma \and
Jingli Xu \and
Linhao Meng \and
Bingru Lin \and
Ying Xu \and
Xiaonan Luo \and
         Wei Chen 
}


\institute{Noptanit Chotisarn, Junhua Lu, Libinzi Ma, Jingli Xu, Linhao Meng, Bingru Lin, Ying Xu, Wei Chen \at
              State Key Lab of CAD\&CG, Zhejiang University, Hangzhou, China \\
              \email{chotisarn@zju.edu.cn, akiori@zju.edu.cn, 504321494@qq.com, 1220341948@qq.com, alice.menglh@gmail.com, linbingru@zju.edu.cn, xu\_ying1027@163.com, chenvis@zju.edu.cn} \\
              Wei Chen is the corresponding author.
           \and
           Xiaonan Luo \at
              Guilin University of Electronic Technology, Guilin, China \\
              \email{luoxn@guet.edu.cn}
}

\date{Received: date / Accepted: date}

\maketitle
\begin{abstract}
Hashtag data are common and easy to acquire. Thus, they are widely used in studies and visual data storytelling. For example, a recent story by China Central Television Europe (CCTV Europe) depicts Brexit as a hashtag movement displayed on an animated bubble chart. However, creating such a story is usually laborious and tedious, because narrators have to switch between different tools and discuss with different collaborators. To reduce the burden, we develop a prototype system to help explore the bubbles' movement by automatically inserting animations connected to the storytelling of the video creators and the interaction of viewers to those videos. We demonstrate the usability of our method through both use cases and a semi-structured user study.

\end{abstract}

\section{Introduction}
Hashtags are an essential part of Twitter trends. Using Twitter's hashtags for Internet activism is called ``Hashtag activism''~\citep{carr2012hashtag}, commonly used to make people aware of social and political issues. Hashtag activism occurs when vast amounts of postings appear under the usual hashtagged words, expressions, or paragraphs in social or political cases via social media \citep{yang2016narrative}. This type of information can be easily acquired through web crawling techniques and is widely used for scientific researches (e.g.,~\citep{sun2017socialwave,wu2018streamexplorer}) and data storytelling (e.g., data journalism).  

Recently, Brexit has been discussed on online social media all over the world. A data story presented by CCTV Europe shows the dynamic of hashtags on animated bubble charts and has attracted much attention in China. This presentation is inspired by the famous animated visualization of Hans Rosling's talk\footnote{https://www.youtube.com/watch?v=jbkSRLYSojo}. The movement of the bubbles (hashtags) are interrelated and can illustrate interesting events to tell stories. As a collaborator of this Brexit visualization story project, we provided various types of visualization, including an animated bubble chart, to support the visual presentation of the data. However, the creation of the final story requires multiple additional steps. The data journalists should first understand all the data and write news scripts based on specific themes. Subsequently, they need to explore the data and find stories with the provided visualizations based on the scripts. Finally, proper animations (e.g., highlighting, slow-motion) are added based on the previous findings to enhance the animated visualizations. This process requires much manual work, and journalists have to switch off between different tools or discuss with collaborators to achieve the final results. 

To reduce efforts during the creating process and provide more opportunities for data storytelling, we present Brexble (stands for \textbf{Bre}xit bub\textbf{ble}), a storytelling prototype system with animated data visualization. This system supports automatic data storytelling through animated bubble charts. We use scatterplot for visualizing data, as it is intuitive to read~\citep{xie2018semantic} and can help gain a good understanding of its distribution. Then we analyze the data to determine the level of movement of each hashtag. The intense level was relevant to essential events. Therefore, we can map the movements to animations to emphasize the events that allow the hashtags to tell a story. The story can be enhanced by fine-tuning animations and adding captions. The narrators can choose to outline the event by utilizing all the provided hashtags, or they can select hashtags from the system. The viewers can watch the videos created by the narrators via the system. To validate the system's usefulness, we conducted semi-structured interviews to measure the awareness of stories by viewers and narrators. Awareness is the extent to which  the viewers and narrators can perceive the relationship between the animated bubble movements and the events of the story through the use of captions. The system is available online at \href{https://bit.ly/2SHqEjw}{https://bit.ly/2SHqEjw}.


To summarize, the contributions are twofold.
\begin{compactitem}
\item A method that defines the three types of animations used for emphasizing the important parts of storytelling through the movement of bubbles,
\item A prototype system that provides automated animations to support the storytelling of bubble movements, which represent real-life events.  
\end{compactitem}

\section{Related work} 
\subsection{Data Journalism} 
The phrase ``data journalism'' offers new possibilities to combine the traditional ``nose for news'' and the ability to tell a compelling story with the sheer scale and range of available digital information by using programming to automate the process of gathering and combining data from various sources \citep{gray2012data}, including online social media sources or other digital media platforms. As stated by Lorenz \citep{lorenz2010data}, data-driven journalism is primarily a workflow that consists of the following elements: digging deep into data by scraping, cleansing and structuring, filtering by mining for specifics, visualizing, and making a story. One way to tell an engaging story is a data video that incorporates visualizations about facts. This method has become increasingly popular as a means of telling stories through data \citep{amini2015understanding}, and Young \etal suggested to tackle data by using techniques that encourage a mixture of exploration and explanation \citep{young2018makes}.

Our authoring tool provides two abilities of visual data-driven stories, which are to communicate a narrative and present information based on data via video playback by following the data journalism pipeline.

\subsection{Visualization authoring tools}




A survey \citep{young2018makes} indicated that the interactions most used by journalists are straightforward data techniques, such as, ``inspect'', ``filter'', ``extract'' and ``elaborate''. A study on systematically identifying factors by proposing a design space derived from the reading experiences in 80 data-driven visual storytelling, which is presented in seven factors, Navigation input, Level of control, Navigation progress, Story layout, Role of visualization, Story progression, and Navigation feedback \citep{mckenna2017visual}.

There are some programming-based tools for authoring visualizations \citep{li2018echarts,mei2018viscomposer}, but they require much programming skills and visualization expertise. Some interactive approaches are more usable for casual users. ChartAccent \citep{ren2017chartaccent} allows charts to be added quickly and easily through a series of interactive notes that create self-explanatory and data-driven annotations. 
VisJockey is a technique that enables viewers to easily access the author's intended view through Orchestrated Interactive Visualization  \citep{kwon2014visjockey}. Some visualization forms of VisJockey requires the user to keep scrolling to read the content continuously. DataClips \citep{amini2016authoring} allows non-experts to collect data-driven ``clips'' to create longer sequences and export as a video clip with added captions. iStoryline~\citep{tang2018istoryline} is a tool that integrates high-level user interactions into automatic layout optimization algorithms to balance manual and automatic storyline layouts.




The sharing function in data journalism is also an essential part for journalists to increase audience reach in news and discussions, indicating the quality of their work \citep{young2018makes}. In the Gapminder\footnote{https://www.gapminder.org/tools/} authoring tool, there is a sharing function that can share/embed the HTML \emph{iframe} for websites.

We studied the surveys mentioned and compared them to the quoted authoring tools to create the design space. The narrator can tell multiple perspectives of the story from the dataset. Furthermore, video playback can support sharing to increase viewer awareness. We present the data movement through automatic animation and share it via a video format that viewers can instantly receive from narrators through video playback. 
\section{Data}
\begin{figure}[htp]
    \centering
    \includegraphics[width=\linewidth]{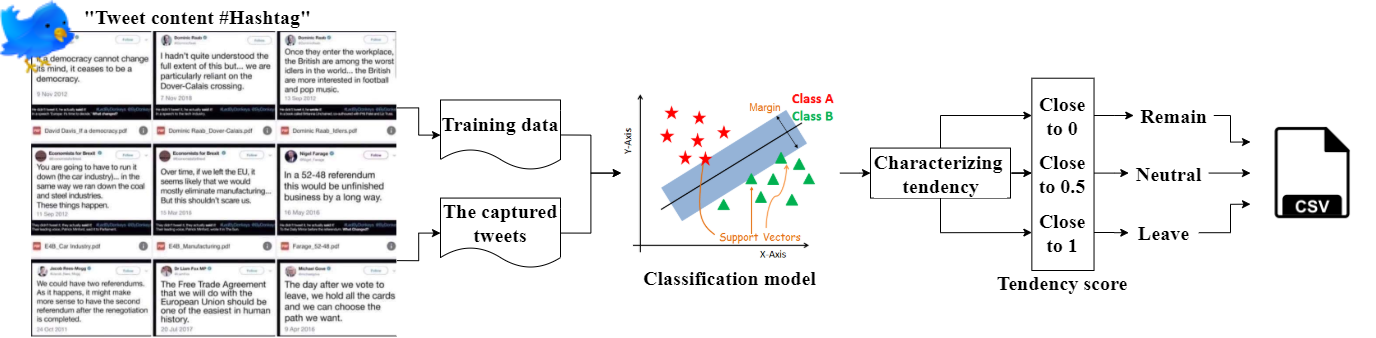}
    \caption{The following steps determine how the data preparation team processed the data: firstly, the Brexit tweets were collected from twitter. Secondly, the tweets were tagged as ``leave'' and ``remain'' side for training data. Thirdly, the SVM classification model was run to categorize all the captured tweets by polarization. Next, the tendency scores were from the probability. Finally, a CSV file was summarized for finding insights.}
    \label{fig:datapipeline_1}
\end{figure}

In this section, we describe the Brexit dataset used to create the visualization. This dataset is generated based on the requirements of the journalists from CCTV Europe. We provide an overview of the Brexit dataset's data preparation process in Section \ref{Data characteristics} and Section \ref{Data preprocessing} to explain how the data preparation team processed the data for visualization (Figure \ref{fig:datapipeline_1}). They collected data from twitter, tagging ``leave'' or ``remain'' side for training data, run the classification model to categorize all the captured tweets into ``leave'' vs. ``remain'' and to determine the size of the bubble and polarization by tendency score. For other datasets, these processes may not be used to prepare the data.

Section \ref{Finding insights in data} explains how to find insights and converts the data obtained from Section \ref{Data preprocessing} into story movement data used to tell the story further. Another dataset, COVID-19 dataset, has been prepared without passing the preparation process described in Section \ref{Data characteristics} and~\ref{Data preprocessing}. The COVID-19 data is readily available from the European Center for Disease Prevention and Control website\footnote{https://www.ecdc.europa.eu/en/publications-data/download-todays-data-geographic-distribution-covid-19-cases-worldwide}.

\subsection{Brexit data characteristics} \label{Data characteristics}



The characteristics of the Brexit dataset applied to the visualization is a kind of political polarization, which is a tendency to participate in politics on one side of the left-right political spectrum \citep{garimella2017long}, that is currently available online in the form of internet activism on Twitter. The Brexit topic consists of bipolar opinions between leaving the EU and staying in the EU. The middle ground between these two is considered to be neutral. In the following part, we explain the rationale of dividing Brexit datasets into three viewpoints: leave, remain, and neutral.


\subsection{Brexit data preprocessing} \label{Data preprocessing}



This section will explain the output of the tendency value $x$ to categorize selected hashtags within the political polarization of the Brexit topic. $x$ ranges from $0$ to $1$, where $1$ represents a complete leave, and $0$ represents a complete remain. The closer to $1$, the more obvious the tendency to leave the EU. The closer to $0$, the more obvious the tendency to remain in the EU. Those close to $0.5$ are considered as neutral.

\subsubsection{Collection and tagging hashtags data}
Twitter Developer API was used to capture 41 monthly tweets from January 2016 to May 2019, by first gathering tweets under the ``trending” categories of \emph{\#brexit} and \emph{\#referendum}. Subsequently, the monthly tweets data was captured according to the required hashtag. 

Tweets were labeled as ``remain'' or ``leave'' and used as training data for the classification model.  Some of these hashtags are very emotional, meaning there is no need to judge anything else to comprehend the tweet trend. For example, the tweets with the hashtag \emph{\#voteleave} almost certainly referred to Brexit.


We pre-marked tweets with the following topics as leave: \emph{\#voteleave, \#marchtoleave, \#takecontrol, \#leaveeu, \#standup4brexit, \#no2eu, \#nodeal} and tweets with the following topics as remain: \emph{\#voteremain, \#peoplesvotemarch, \#bremain, \#remainernow, \#abtv, \#yeseu, \#strongerin}.



\subsubsection{Building the classification model}
The SVM classification model was used to categorize all the captured tweets into ``leave'' vs. ``remain'' dichotomies. First, it is necessary to clean the initial tweets with hyperlinks, non-English characters, and then evenly format them into two phrases with space. Lastly, convert them into English lowercase. 

TF-IDF is used to extract feature information. The token pattern of the word is expressed by the regular expression, which means the beginning of the letter or \#, followed by one or more non-null characters until the word boundary. Finally, the positive and negative data marked by Vector-quantization are inputted into the SVM classification model to identify tweets inclined to leave or remain in the EU. 


\subsubsection{Characterizing tendency}
Na\"ive Bayes is a probabilistic model that weighs the probability of a given classification under a given condition to reveal the probability of each feature of the specified class. In this Brexit dataset, Polynomial Bayes is adopted, whose parameters are set without learning prior probability. Additionally, all the data previously classified as the data of known labels are re-used. Finally, the tendency score corresponding to each hashtag is generated. The tendency score will be used to determine the size of the bubble and determine the polarization mentioned at the beginning of Section \ref{Data preprocessing}.

\begin{figure}[htp]
    \centering
    \includegraphics[width=\linewidth]{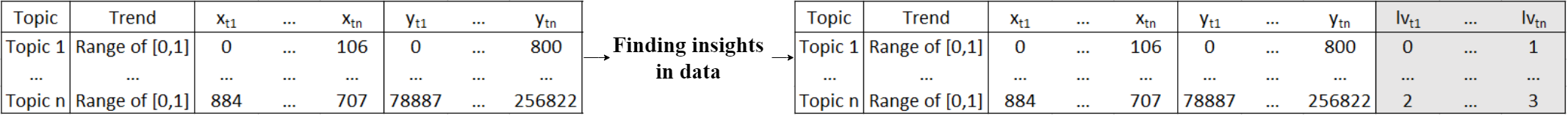}
    \caption{This figure shows the structure of the CSV files, before and after mapping the levels of movement, where the ``Topic'' column has a list of hashtags, ``Trend'' columns showing tendency scores. The ``x'' and ``y'' columns refer to the number of tweets and retweets. Furthermore, the ``lv'' columns mean the level of x and y over time.}
    \label{fig:datapipeline}
\end{figure}

The result is in a CSV file (Figure \ref{fig:datapipeline}) with a structure in which the tendency score will be taken into the ``Trend'' column to determine the size and polarization of bubbles. The data in the ``x'' and ``y'' columns represent the total number of tweets and retweets with those hashtags (``Topic'' column). The data will be used to find the data movement (the grey area) through processes in Section \ref{Finding insights in data}. As mentioned before, Section \ref{Finding insights in data} can be applied with other input data, such as the sample COVID-19 that we present, and will be transferred to the animation mapping pipeline in Section \ref{Animation mapping pipeline}.

\subsection{Finding insights in data} \label{Finding insights in data}
This step aims to start with the data and finish with a story \citep{gray2012data}. The datasets are used to visualize images so that they are easy to analyze and interpret. We use scatter-plot prototypes for visualizing data and gaining a better understanding of its distribution. Then we analyze and understand the data by using the clustering analysis to determine the importance of the event. Notes on the insights are taken to check back and used to transform the data by labeling it into four levels to specify the level of bubble movement, and to identify the importance of the event at a specific time with a line chart. We propose an animation mapping pipeline for telling data stories from finding insight into the preprocessed data.





\subsubsection{Clustering analysis}


This project intends to specify the level of bubble movement and identify the event's importance at a specific time. The narrator uses bubble movement to tell the story through essential facts that connect each topic.

A cluster analysis was used to determine the importance of hashtag movements by plotting the dots of all hashtags, which equates to 36 hashtags within 41 periods, a total of 1476 dots on the scatter plot. We determine the threshold of this dataset more than an average of all tweets (98) because hashtags occur in tweets first, followed by retweets. With this condition, the initial 1476 dots decreased to 282, and the dots that have tweets lower than the threshold are grouped as the Zero level. The 282 dots were applied with the clustering algorithms and compared with the Python library sklearn. 

Clustering algorithms, \eg $k$-means, Affinity Propagation, Mean Shift, Spectral Clustering, Agglomerative Clustering, DBSCAN, and HDBSCAN were examined on the dataset. It was discovered that $k$-means is suitable. 
The elbow technique \citep{kodinariya2013review} is a helpful graphical instrument to estimate the ideal amount of clusters for a specified assignment $k$. The concept behind the elbow technique is to quickly define the value of $k$ where the distortion starts to decline, which will become more apparent if the distortion for distinct is plotted $k$. Based on the plot result, the elbow is at $k = 3$, which indicates that $k = 3$ is a proper choice for this dataset.


We apply $k$-means clustering to determine ``natural'' groupings of instances for a given unlabeled data in predefined classes. After this step, the 282 dots can be separated into three classes, namely, Low level, Medium level, and High level. When combined with the Zero level mentioned above, there are a total of four levels. The automatic is used to reduce the burden of authoring. In the future, we can use a more curated method (e.g.,~\citep{weng2018srvis}) to allow narrators to assign the importance of events interactively.

\subsubsection{Transform data by labelling four movement levels}

As a result of the clustering analysis from the previous insights process, we have determined the bubble movement level. It will be divided into four levels; the first level is very shallow, level 0, the second group is low, level 1, the third group is medium, level 2, and the fourth group is high, level 3.

The four levels of each hashtag can be used to draw a line chart where the x-axis has 41 periods, and the y-axis has four levels. Each hashtag will have its line chart bringing the total to 36 line charts. The line charts are called hashtag movement behavior. It will be included as a result of the Hashtag Pulse presentation (Section \ref{Hashtag pulse}).



From the dataset, 41 monthly periods represent essential events, but the effect it has on each hashtag is not the same. The tipping point of the line chart means a precise movement that may be affected by relevant events at that time. It means that each change interacts differently with the main events. Therefore, we can map the events with the tipping point, allowing the hashtags to tell a story.

\begin{figure*}[htp]
    \centering
    \includegraphics[width=\linewidth]{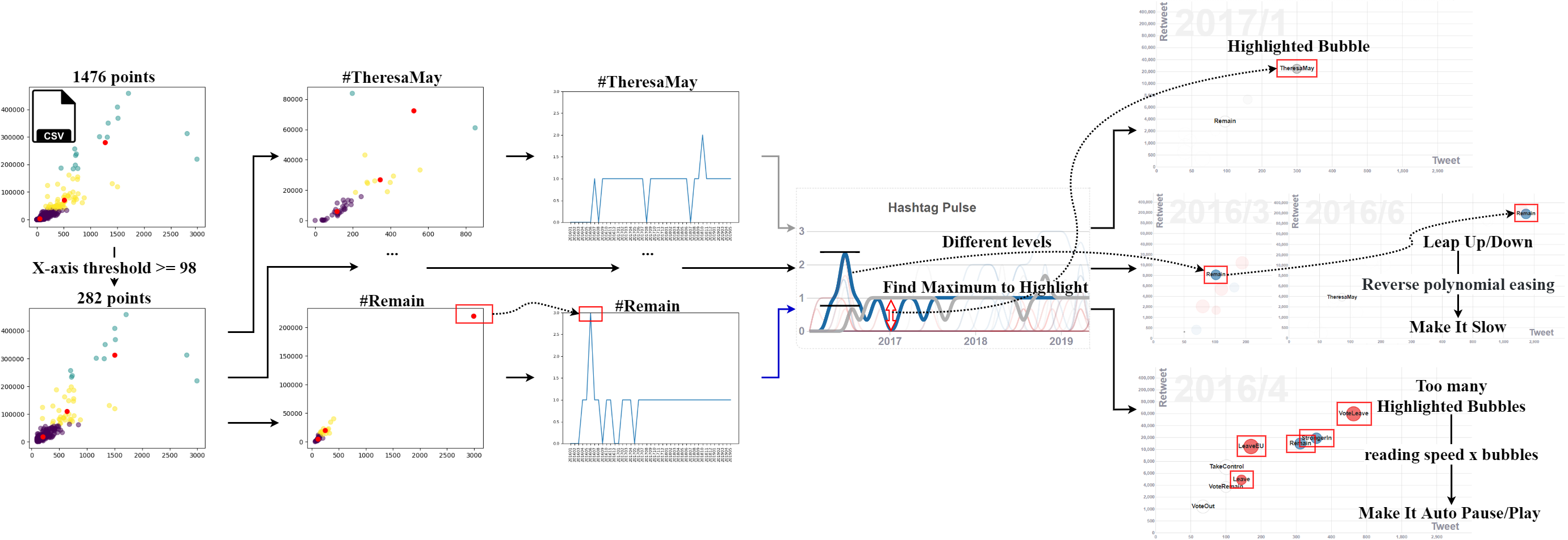}
    \caption{The following steps are required to carry out animation mapping: firstly, the data is extracted and plotted. Secondly, the data is filtered and grouped into four levels. Thirdly, the level of each hashtag is transformed into a movement line, and lastly, the level of movements is compared to define the three animations.}
    \label{fig:systemabtraction}
\end{figure*}

\subsubsection{Animation mapping pipeline} \label{Animation mapping pipeline}


This part summarizes the process mentioned above for the animation mapping pipeline (Figure \ref{fig:systemabtraction}). In the first step, 36 hashtags are imported and multiplied by 41 months, which equates to 1476 times. Hashtags with tweets below threshold (98) are considered to have the lowest movement level. For the second step, only the hashtags with more than the threshold were selected, which leaves only 282 dots. The 282 dots were divided into three levels using $k$-means. As for the third step, the level of each hashtag is transformed into a movement line. The fourth step compares the movement level of each hashtag, which resulted in various animations as follows; the highest value of each period used to determine hashtag highlighting—the value change between two periods used to determine slow-motion. Moreover, when there are many highlighted hashtags, it was applied to determine the pausing to let the user be aware of the highlighted hashtags. From this pipeline, it can be truncated by importing other inputs, such as COVID-19 dataset, to find the level of motion of the bubbles for further visualization.

\subsection{Brexit data stories example}
\begin{figure}[htp]
    \centering
    \includegraphics[width=\linewidth]{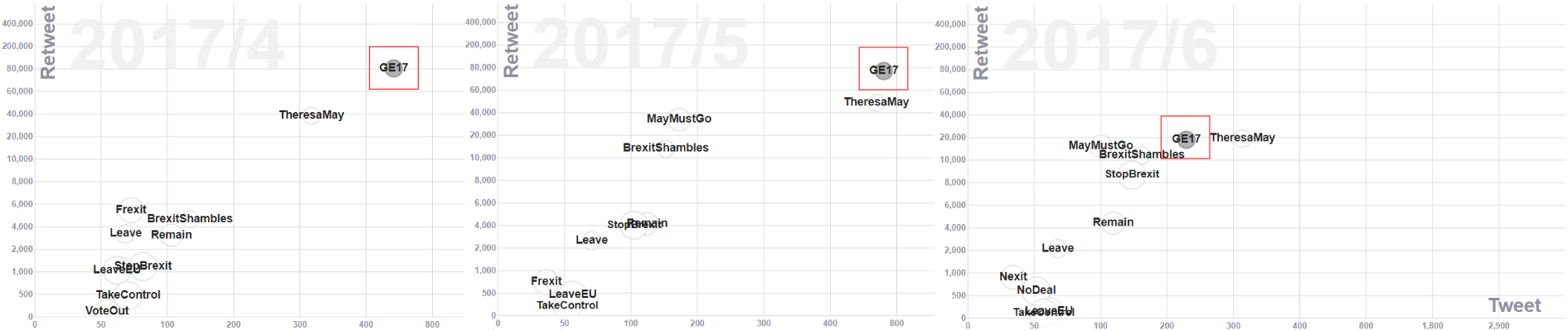}
    \caption{\emph{\#GE17} (highlighted with a red rectangle) moves at a more intense level than all other hashtags from April to June 2017. }
    \label{fig:GE1704-06}
\end{figure}

An example of a movement that can classify events is the \emph{\#GE17} moving at a more intense level than all other hashtags from April to June 2017 (Figure \ref{fig:GE1704-06}). The movement corresponds to the critical events of that period: Prime Minister calls a General Election - to be held on 8 June 2017.

It can be seen that if we do not emphasize the \emph{\#GE17} bubble, we may neglect this important election event. So we offer a way to understand the movement of these hashtags by focusing on the movement level. To reiterate, the movement level of each hashtag corresponds to actual events. Therefore, narrators must first understand the real story and then choose a viewpoint to tell the story through the movement of hashtags.

\section{Visualization design}

Brexble has been created through an iterative customization process of evolutionary prototyping. The first version of the system was edited via post-production to present the news on CCTV Europe. It was extracted and compared with the next version that has an automated animation movement function to see whether or not the movement is consistent with the first version. It was discovered that a movement could tell the story by itself; the function of captioning was developed so that narrators could create captions while also choosing the hashtags they want to present. Subsequently, it can be exported as a video for viewers.

\subsection{Design rationales}
Brexble visualization allows the narrators to create their story and emphasize the bubble movements by using the captions that need to convey the story in the form of captions from different points of view based on these design tasks, namely, exploration, explanation \citep{riche2018data}, and suggestion. 

The exploration task provides narrators with the power to find their own story in a set of data. In this study, the bubble chart can display the color and size of the bubbles, indicating the side and degree of inclination to either side of the Brexit campaign. Narrators can choose and explore the bubbles they want to convey in the story.

While the explanation task communicates the narrators' story from the data, storytelling is the ability to explain stories. Narrators can choose which topic to tell via the selected bubble. Moreover, storytelling is the ability to add captions. The captions used here explain the movement of the bubbles at the desired time and are presented in the form of a short message subtitle. 

The suggestion task is also added to help the narrator create their story. The bubbles have their movement level, based on the recommended automatic animation, which is either highlight, pause, or slow-motion. Highlights are used for emphasis. Automatic pauses are also used so the audience could understand better. Slow-motions are also included so the receiver can notice the crucial.

\begin{figure}[htp]
    \centering
    \includegraphics[width=\linewidth]{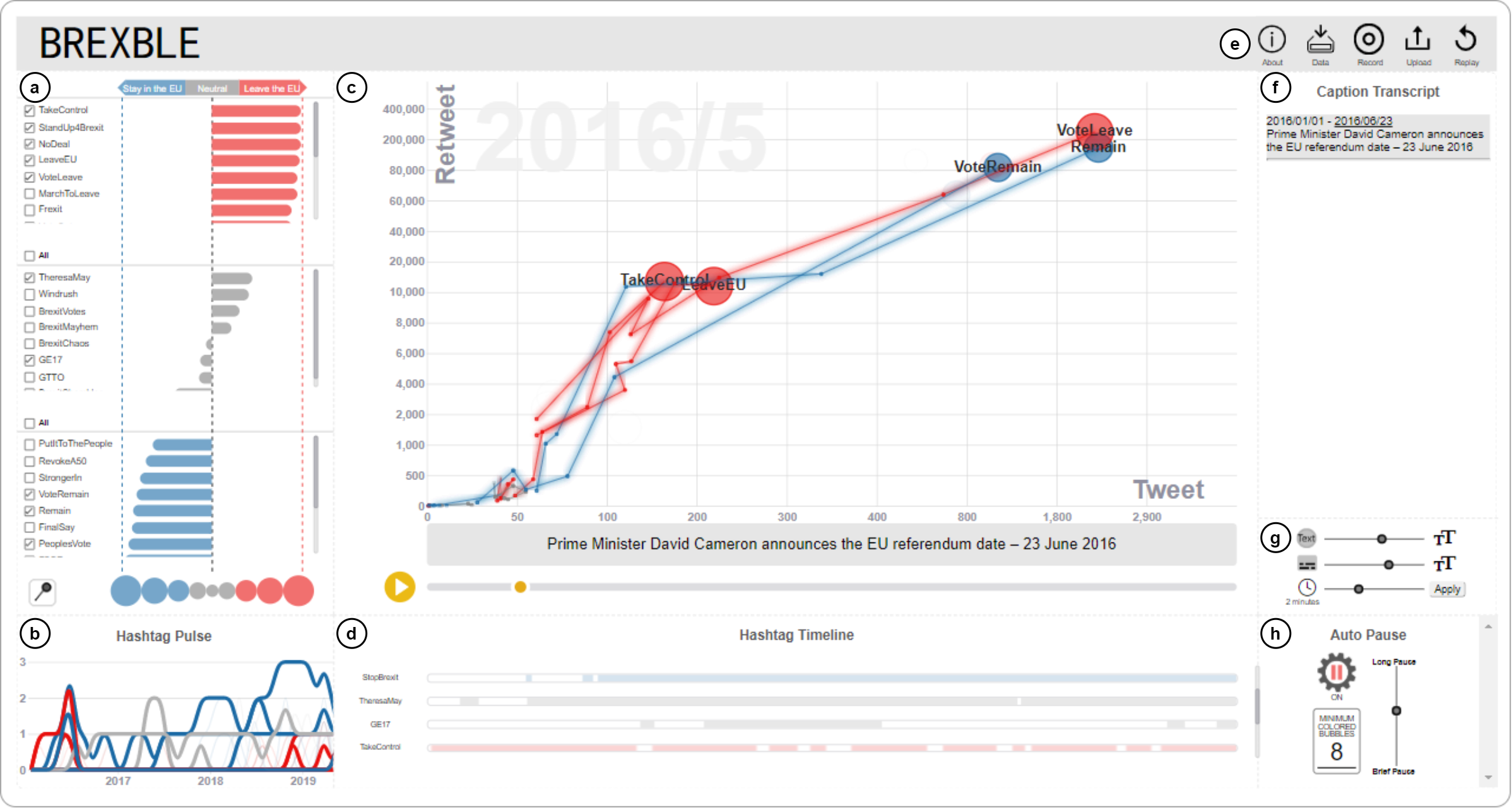}
    \caption{A browser-based visual Brexit storytelling authoring tool, Brexble, portraying a timeline of one month before the EU referendum, where each color corresponds to either political polarization; the interactive bubble chart can be accessed at \href{https://bit.ly/2SHqEjw}{https://bit.ly/2SHqEjw}. A demonstration video can be viewed at \href{https://youtu.be/huuko8p3-e4}{https://youtu.be/huuko8p3-e4}.}
    \label{fig:teaser-1}
\end{figure}

\subsection{User interface}
The main elements of the hashtag storytelling are described as follows. 

Bubbles are the key visual elements of the visualization. The \textbf{colors} encode the polarization: blue represents the side that wants to stay in the EU, red represents those want to leave, and grey is used for neutral bubbles (Figure \ref{fig:teaser-1}(a)). Bubbles that have not yet been specified and given a color remain white until it has been highlighted.
The \textbf{size} of the bubble is the extent of the hashtag's inclination toward either side (Figure \ref{fig:teaser-1}(a)), whereby the bigger it is, the more inclined it is. For the neutral bubbles, since there is not much inclination, the bubbles are never significant.
The \textbf{trajectories} of the hashtags show the direction and trend of each bubble's movement. 

\subsubsection{Exploration and editing panel}
\textbf{Hashtag bubble selection} at the top-left of the system (Figure \ref{fig:teaser-1}(a)) is where narrators can choose which bubbles they want to tell their story. By default, every bubble will be shown. However, if the narrator chooses only some bubbles, those bubbles will be displayed with highlights automatically shown. The narrator can turn the function that displays the hashtags'  \textbf{trajectory} on or off at the bottom of the Hashtag Bubble Selection panel (Figure \ref{fig:teaser-1}(a)).

The \textbf{bubble chart} placed in the middle (Figure \ref{fig:teaser-1}(c)), and the narrator can select the one that matches their story. Under the bubble chart, the narrator can insert captions to explain the movements of bubbles. The video progress bar indicates the timeline of the movement. The main screen and captions can be exported into a video playback.

\textbf{Hashtag pulse} \label{Hashtag pulse}
is a line chart (Figure \ref{fig:teaser-1}(b)) that combines the movement level of every hashtag, which is divided into four levels, 0-3. Hashtags with the lowest level move on the lower level and may rise to the middle of the chart or go up to the top right corner of the chart when its level of movement changes to the highest level. The Hashtag Bubble Selection panel can be used to display information on the selected hashtag. 
Therefore, the narrator can compare hashtags.

\textbf{Hashtag timeline} (Figure \ref{fig:teaser-1}(d)) is used with Hashtag Pulse. 
It helps narrators carry out analysis and make informed decisions.  Each hashtag has a different timeline of occurrence, existence, and disappearance. The system calculates the occurrence in all the selected hashtag timelines. Whichever one occurs first, the first occurrence of that hashtag will be used as the base in the main timeline. The main timeline will affect the start and end of the video progress bar.

\subsubsection{Configuration panel}
In terms of customization, narrators can customize the caption, such as the font size of captions, size of texts on the bubble, video duration, and the beginning and end of the pause of the bubble movement, and record and replay the bubble chart.

\textbf{Caption transcript}:~
In the system, stories can be used by the narrator to describe hashtag movements with a maximum of 160 characters. The narrator can choose which time to tell the story by adjusting the timeline or pressing the play button and waiting until the precise time then fills in the text in the caption box. A maximum subtitle length of two lines is recommended\footnote{https://bbc.github.io/subtitle-guidelines/}. While filling in the text, the bubble chart will not move until the narrator clicks the play button again. Narrators can edit the caption on the Caption Transcript panel on the right (Figure \ref{fig:teaser-1}(f)). They can customize the start and end time for each caption by clicking on the date, which can be deleted by double-clicking on the text. 
In the bottom of the Caption Transcript panel, narrators can customize the caption, the font size of the caption, size of texts on a bubble, video duration (Figure \ref{fig:teaser-1}(g)).

\textbf{Auto pause}:~
Pausing at Auto Pause panel (Figure \ref{fig:teaser-1}(h)) is the only animation that narrators can choose to enable or disable animation displays. The narrator can specify the minimum number of colored bubbles moving during the same period to pause when they meet certain conditions. This function enables the narrator to observe and discern the color of bubbles that are being displayed at that time. Narrators can also specify the duration of the pause by calculating the amount of time a human can read an object multiplied by the number of bubbles.

\textbf{Record and replay the video}:~
Once the narrator has written a caption to describe the hashtag movement of the bubble and adequately adjusted the size of the characters and the pausing duration, the narrator can record the video using the top bar (Figure \ref{fig:teaser-1}(e)). While recording, the narrator can continue to interact with the bubbles on the screen, such as scrolling the mouse to adjust the size of the text on the bubble. The interaction will be recorded as a JSON file for uploading and replaying later in video format.

\subsubsection{Animation}
Three animations were created from the animation mapping pipeline (Figure \ref{fig:systemabtraction}) and automatically added to the bubble chart to give the story clarity and meaning. 

\textbf{Highlight}
As mentioned before, during each period, there are different hashtags with different levels of movement, which are identified by highlights. This allows the narrator to tell stories based on the highlighted hashtags, which signify the importance of real events. Highlighting allows the bubble with the highest movement levels to remain and be shown in its original color, while the bubbles that are not the highest value will be in white. In the case where all the bubbles are displayed without any particular selection, the narrator can enable or disable the display of the bubbles not highlighted.

\textbf{Pause}
Although the essential bubbles have been highlighted, it may still be difficult for the narrator to focus on each highlighted bubble simultaneously. For this reason, We offer automatic animations that can temporarily pause the movement of the bubbles at locations with many highlighted bubbles. The duration of the pause is calculated from the speed reading per word per number of highlighted bubbles, which is determined by the narrator or in the system by default. Once pausing has expired, the bubbles will continue to move along the timeline.

\textbf{Slow-motion}
The movement level of bubbles used to calculate and adjust the speed to a slow-motion will be taken from the highest level of bubble movement in each period. Different movement times have different movement levels, whereby the higher level of movement, the faster the bubble moves. Moreover, when the bubbles move from one time to another, it will compare the movement values of those two periods, and if the maximum values of the two periods are higher, the bubbles will move faster. Nevertheless, this will be adjusted accordingly, whereby the higher the value, the most slowly the bubbles will move. This decrease in speed will be done by employing the \verb"d3.easePolyOut(t)" function where \verb"(t)" refers to the difference between the maximum values of the two time periods. 
\subsection{Interaction}

The provided interactions are grouped according to the categorization of interaction techniques proposed in \citep{yi2007toward}. The narrator can tell the story from the hashtag movement behavior through the following interactions;



\textbf{Selection \& Exploration} can help the narrator to explore the display, movement behavior, life span of the topic that can bring to tell the story. By default, Brexble displays all the hashtags on the bubble chart. Brexble allows narrators to select interest hashtags that displayed prominently from the Hashtag Bubble Selection panels, and the chosen hashtag will display on the bubble chart, the Hashtag Pulse, Hashtag Timeline.


\textbf{Reconfiguration} enables narrators to choose their interested topic to further tell a story by comparing the animation of the chosen topics. In every hashtags selection, the bubble presentation will have different automatic animation insertions, so the movement of each selected bubbles being displayed by the arrangement of highlighting, pausing, and slow-motion in a different combination.



\textbf{Abstraction \& Elaboration} allow narrators to alter the representation of the animations by bubbles selection. Brexble enables narrators to select the bubble, which will automatically show the highlighted bubbles. It is related to pausing, which can be activated when many highlighted bubbles are displayed at the same time. Pausing can adjust the minimum number of bubbles and the pause duration. Slow-motion can be auto-adjusted when the selected bubble has changed and represents a leap motion.


\textbf{Filtering} enables narrators to change the items being presented based on some specific conditions or some critical information. They can select all the hashtags to show, they are allowed to view only the highlighted bubbles. Thus, the non-highlighted bubbles will make the highlighted ones more noticeable.

\subsection{Prototype implementation}

The back end of the pipeline that provides the data handling for the extract and format tasks is implemented in Python. The front end that supports the presentation of the data is implemented with D3.js. We use rrweb.io\footnote{https://www.rrweb.io/} to export a JSON file of the recorded screen of the bubble chart, which can be replayed as a video through the Brexble system. 
\section{Evaluation}

This section discusses the design user study of the system through the use of the Brexit hashtag activism case study. The case study is divided into four types of user study, which will be summarized at the end. We have also applied Brexble to a dataset that captures COVID-19 epidemic in Southeast Asia.

\subsection{Brexit hashtag activism case study}

Bojo is interested in the news of the Brexit and wants to create a video based on Brexit events through the movement of hashtag bubbles. Initially, he clicks on the play button to see all the movements of the bubbles from start to end with the default animation of the system. He explores the overall movement level by using Hashtag Pulse, together with Hashtag Timeline.

Then, he selects a few hashtags and presses play to view the movement of those hashtags. The animation conditions will display different results depending on the selection of the hashtags. At this point, Bojo writes a caption to explain the events and hashtag movements to tell the story in his chosen style. If Bojo has not chosen any hashtags, the system will display all the bubbles by default. All the colored hashtags with the default number of four are displayed at the same time. The movement of the bubbles will pause, and Bojo inputs the caption at that time.

The critical pause periods are at the 2$^{\textrm{nd}}$, 4$^{\textrm{th}}$, and 5$^{\textrm{th}}$ month, which are major hashtag events on both the leave and remain sides and are highly contested because this period is before the referendum on the withdrawal from the EU. Furthermore, after the referendum, the bubbles that have been competing on both sides reduced the level of competition between June and July 2016, which was a very remarkable event and resulted in a slower system.

Subsequently, the moment Theresa May became the Prime Minister was considered to be a standout event that is of great importance, and so captions have been added. After that, a new election was held in 2017, which is also seen as a prominent time and so captions are also added. After the 2016 referendum, numerous calls for the revision of the referendum results were requested, which Bojo can see represented in the hashtag presentation. Therefore, he wrote a caption describing the story here as well.

This user story demonstrates that highlights, pauses, and slow-motion added to the hashtag bubbles can allow the narrator to communicate a variety of stories. As a result, by experimenting with narrators and viewers, four types of storytelling were defined, which will be discussed in the next section.

\subsection{User study}


This visualization displayed the changes that occur over time \citep{andrienko2003exploratory}: 

\begin{enumerate}[topsep=0pt,itemsep=-1ex,partopsep=1ex,parsep=1ex]
\item Existential changes, 
\item Changes of spatial properties,
\item Changes of thematic properties expressed through values of attributes.
\end{enumerate}
We consider the movement of the bubbles in the space over time to consist of three co-dependent parts: ``Where'' (location/space), ``When'' (time),``What'' (objects) \citep{peuquet1994s}.

Hashtag activism is presented along with the time-series data, meaning that each event should be mapped with specific periods. A Brexit timeline \citep{walker2018brexit}, provided by the House of Commons Library, discusses the events leading up to the UK's exit from the EU. The users are offered a summary of the important events from the Brexit timeline as an awareness guide on the movement of the bubbles.

\begin{figure}[htp]
    \centering
    \includegraphics[width=.50\linewidth]{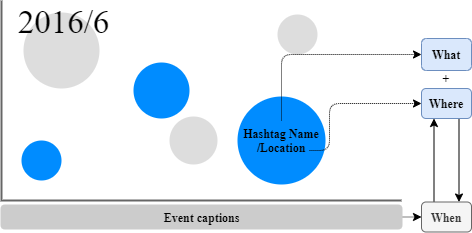}
    \caption{Three co-dependent parts show the relations of ``What'' + ``Where'' and ``When''.}
    \label{fig:3W}
\end{figure}

The important moments (``When'') from the Brexit timeline or the one created by the narrators will be used to describe the bubble chart to see the relationship between the movement of the bubbles and the captions. Movement refers to the position in space, also known as location, the object is ``What'', and the location is ``Where''. The relationship between three co-dependent parts outlines in Figure \ref{fig:3W}.

Both sides of the experiments, the narrators and the viewers have been designed. The narrators will be required to create a video of the Brexit timeline story through events (``When'') and specify and describe the ``What'' and ``Where''. A semi-structured interview was carried out on both the narrators and viewers, which discusses the ``What'' and ``Where'' questions of hashtags and included a part where they were asked to identify and explain what was perceived from the videos.

On the narrator's side, two narrators were asked to create videos. One is an IT researcher for social research and created the first and second video. The other, a social media researcher, created the third and fourth video.


On the viewer's side, five IT students, who were interested in Brexit, were asked to participate in a semi-structured interview. They watched the video that the narrators created. They were asked to specify their awareness of the bubble chart movement and how it relates to the events on a scale from 1-5 (1-considered ``not at all aware'' and 5-``extremely aware''). Moreover, viewers were asked to share an overview of the events that occurred with the movement of the bubbles.


For this research, four videos were created based on ``When'' and ``What'' + ``Where''.

``When'' is divided into two types; 
\begin{enumerate}[topsep=0pt,itemsep=-1ex,partopsep=1ex,parsep=1ex]
\item The researcher specifies the events from the Brexit timeline \citep{walker2018brexit}.
\item The narrator specifies the events.
\end{enumerate}

``What'' + ``Where'' is divided into two types; 
\begin{enumerate}[topsep=0pt,itemsep=-1ex,partopsep=1ex,parsep=1ex]
\item Select all the hashtags that the system provides.
\item Select from some of the hashtags that the system provides.
\end{enumerate}

Subsequently, video versions are created as follows;
\begin{enumerate}[topsep=0pt,itemsep=-1ex,partopsep=1ex,parsep=1ex]
\item Version 1; Specify ``When'' and Specify ``What'' + ``Where''
\item Version 2; Specify ``When'' and Do not specify ``What'' + ``Where''
\item Version 3; Do not specify ``When'' and Specify ``What'' + ``Where'' 
\item Version 4; Do not specify ``When'' and Do not specify ``What'' + ``Where''
\end{enumerate}

The results from the sequence of the videos were reported, where 1 means ``not at all aware'', 2 refers to ``slightly aware'', 3 means ``moderately aware'', 4 indicates ``very aware'' and 5 refers to ``extremely aware''. The interval is 0.8, with the highest score minus the lowest score and then divided by the total number of 5 levels.

\begin{figure}[htp]
\begin{minipage}{.99\textwidth}
    \centering
    \includegraphics[width=\linewidth]{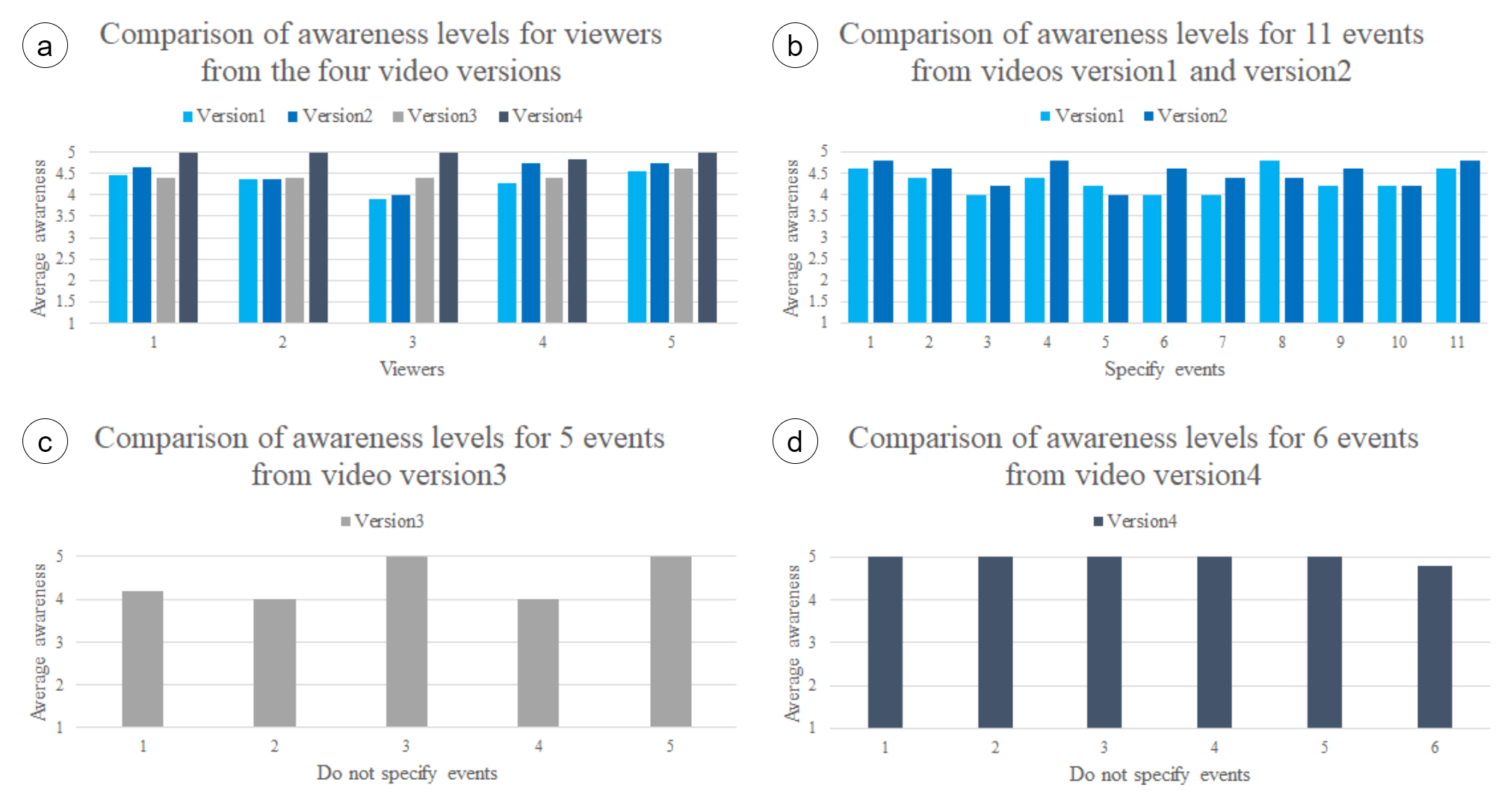}
    \caption{The bar charts show the four versions of Brexit user study result.}
    \label{fig:result}
\end{minipage}
\end{figure}

\textbf{Version 1}: The first narrator created the first video.
This version has been created under the condition that all the provided hashtags must be used with the given events. The events were defined as general events of Brexit. It does not explicitly mention the movement of the hashtag unless the specific word in that event relates directly to the word that was used as a hashtag.

The data results on the awareness of each viewer watching this version (Figure \ref{fig:result}(a)) indicated that four out of five people were extremely aware of the events according to the movement of the hashtags, while one person was very aware.

\textbf{Version 2}: The second video was also created by the first narrator, who was given conditions to use the specified event. In this version, the narrator  chose only the hashtags she knows to tell the story, which are \emph{\#TakeControl, \#LeaveEU, \#TheresaMay, \#GE17, \#PeoplesVote}, and \emph{\#StopBrexit}. The reasons for choosing these hashtags are because the hashtags are consistent and match the keywords in the given event and because they coincide with the personal knowledge of the narrator.

The data results on the awareness of each viewer watching the second version (Figure \ref{fig:result}(a)) portrayed that four out of five people were extremely aware of the events through the movement of the hashtags, while one person was very aware. Results indicated that viewers are more aware of the hashtags used and are also more aware of the slow-motion than the first video. 

The bar chart (Figure \ref{fig:result}(b)) compares the two versions of the video with the control variable, which is an identical event. Data indicated that the viewers watching the second version of the video had a higher level of awareness. During the interview, the audience expressed that they had a higher level of awareness because of fewer hashtags related to the events. However, for events 5 and 8, there was a higher level of awareness for the first version. Moreover, in the final event where many hashtags were presented simultaneously, both the narrators and viewers had more awareness of the hashtags and animations than the first video.


\textbf{Version 3}:
The third version was created by the second narrator, which shows all the hashtags in the system. The narrator specifies the caption by himself, and the movement of the bubble group was emphasized more prominently than the individual bubbles.


The viewers mostly focused on the movement because the captions tell more about bubbles' change than the real Brexit event. The viewers claimed they understand the progress of the bubbles because the captions led them to see its transformation and links with their prior knowledge.

For the third video (Figure \ref{fig:result}(c)), all five of the viewers indicated an awareness level of ``very aware'', and the rest are extremely aware.

\textbf{Version 4}:
The second narrator created the fourth video. In this version, the narrator is free to put in the captions himself and choose which hashtags to tell the story. The narrator chose eight hashtags: \emph{\#VoteLeave, \#LeaveEU, \#Leave, \#VoteRemain, \#TheresaMay, \#GE17, \#Stopbrexit,} and \emph{\#PeoplesVote} because they correspond to his knowledge of the Brexit timeline. The narrator inserted captions that summarize Brexit's importance according to his viewpoint. Moreover, he added an overview of the movement of the hashtags in each event period.

Results indicated that viewers are more aware (Figure \ref{fig:result}(a)) of the hashtags that the narrator chose to convey and are better aware of the slow-motion when compared with other video versions. 

For the events in the fourth video (Figure \ref{fig:result}(d)), almost all of them scored a high level of awareness, apart from one who had a lower score. However, the video is still categorized as obtaining extremely aware results.


\subsection{A Summary of results}

Based on the interviews, the narrators are aware of the highlighted animation, the slowing down when hashtags move too quickly, and the pauses when there are too many important hashtags simultaneously.

As for the awareness level (Figure \ref{fig:result}(a)), results indicated that all video versions scored an awareness level of ``extremely aware'', with an average between 4.21-5.00, which is at the level of 4.31, 4.44, 4.49 and 4.97, respectively.

Most viewers expressed a higher level of awareness in the second and fourth videos than the first and third videos. In the second and fourth videos, the narrators chose the hashtags that they think were important and related to the events, which according to the viewers, made to important events more comfortable to understand because there were fewer hashtags.


\begin{figure}[htp]
    \centering
    \includegraphics[width=.7\linewidth]{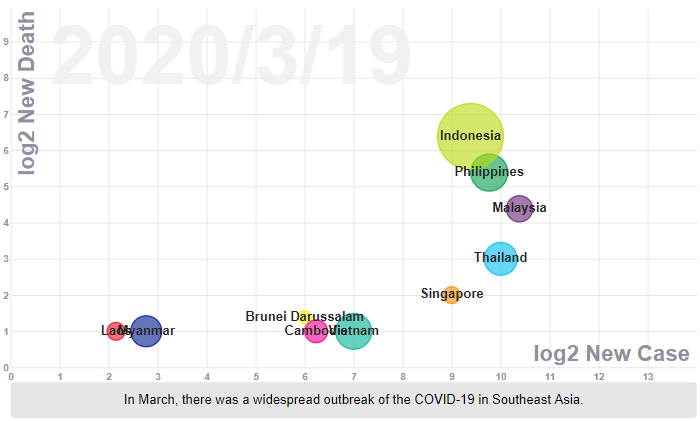}
    \caption{Brexble trial with COVID-19 dataset, named as Brexble $\times$ COVID-19, portraying a timeline of the panic period of Southeast Asia counties under the COVID-19 curtain; the interactive bubble chart can be accessed at \href{https://bit.ly/38aLv5i}{https://bit.ly/38aLv5i}.}
    \label{fig:BrexblexCovid}
\end{figure}

\subsection{COVID-19 case study}
We also applied this visualization to a dataset that reflects the COVID-19 outbreak in Southeast Asia. The sum of weekly new confirmed cases of each country is represented with its x-axis position, and the y-axis position encodes the weekly total number of deaths (Figure \ref{fig:BrexblexCovid}). The bubble size encodes the population of each country.

The topic is changed from the hashtags to Southeast Asia countries without polarization. The number of confirmed cases and the number of deaths can convey the story from a variety of perspectives, \eg public health, foreign affairs, people's anxiety. These perspectives can be chosen by narrators to tell the story using captions of the movement of bubbles.

This dataset is converted to a log scale, which is different from the linear scale in the Brexit dataset. For the time scale, it is different because the COVID-19 dataset is just happening, so it was put into weekly spans, unlike the Brexit dataset, which is divided into monthly spans. For movement levels, $k$-means was still used to divide the movement into four levels.


One case is briefly described as follows. During January 2020, people in the region were not so awakened by COVID-19. However, by the beginning of March 2020, many countries have started to take very intense action, which is consistent with the bubble's automatic highlights during the third month, with highlights being shown in almost every country. By May 2020, it can be seen that many bubbles are not highlighted, which means that many countries are starting to ease up but still lack trust. In June 2020, many countries, especially those with large populations, still have many confirmed cases, and many people continually died. The system is available online at \href{https://bit.ly/38aLv5i}{https://bit.ly/38aLv5i}. 
\section{Discussion}



The narrators used the content from the Brexit timeline to create videos. They tried to determine the duration of the event according to the timeline and observed the hashtag movement to tell a story by selecting specific hashtags from the system. That is, the narrator observed and associated the events (``When'') with hashtag movements (``What'' + ``Where''). On the other hand, the viewers first described the movement of the hashtag (``What'' + ``Where''), then described the event of the movement (``When'') and finally connected the two things and explained their perceptions.




The study included five users as the viewers and two users as the narrators. Nonetheless, an increase in the number of test subjects would be desirable to gather more quantitative data for analysis.

For applying Brexble frameworks to another dataset, we have tried the COVID-19 outbreak in the Southeast Asia dataset that is different from the Brexit dataset in terms of time scale, data scale, or even non-polarization. We found that it was still able to tell its own story from the narrators' perspective. For polarization or grouping topics, it helps the viewer to remember the moving bubbles from colors better than the non-grouping topics.  

So both datasets can tell their own stories through animations and captions, but the narrators may need to have prior knowledge of the topic. Brexble will help the narrators to connect that knowledge in the form of events with the bubbles' movements and then communicate it as a story. 
\section{Conclusion and future work}

The study presents a prototype system to augment the animated bubble chart by automatically inserting animations connected to the storytelling of the video narrators and the interaction of viewers to those videos. It reduces the burden on humans when making visualizations. The proposed prototype system not only helps in exploring changing data pattern but also supports to make conclusions. 

In the future, clustering by employing $k$-means can be adjusted using some other clustering algorithm to label the data for presentation. Furthermore, as data results indicated, most viewers have prior knowledge of the story. Therefore, we should design Belief-Driven Data Journalism \citep{nguyen2019belief}, which is a framework that integrates the viewers’ beliefs into the design to encourage interaction.

\begin{acknowledgements}
This work is supported by National Natural Science Foundation of China (61772456, 61761136020). Moreover, the first author wishes to thank Mr. Jaturapat Patanasongsivilai and Ms. Min Zhu for their valuable technical support on this project. Mr. Jaturapat Patanasongsivilai, the founder of ``facebook.com/programmerthai'' and computer textbook author, for his help to lay the foundation of the three proposed animations. Furthermore, Ms. Min Zhu, a postgraduate student from the Heinz College at Carnegie Mellon University in Pittsburgh, Pennsylvania, United States, for her help in initial collecting the Brexit dataset and consulting about the data preprocessing process.
\end{acknowledgements}


\bibliographystyle{spbasic}      
\bibliography{template}   

%
%

\end{document}